\documentclass[epj]{svjour}
\usepackage{graphics}
\usepackage{longtable,lscape}
\usepackage{rotating}
\usepackage{multirow,multicol}
\usepackage[breaklinks]{hyperref}
\usepackage{natbib}
\begin{document}
\title{Anisotropic Compacts Stars on Paraboloidal Spacetime with Linear Equation of State}
% \subtitle{Do you have a subtitle?\\ If so, write it here}
\author{V. O. Thomas\inst{1} \and D. M. Pandya\inst{2}% etc
% \thanks{\emph{Present address:} Insert the address here if needed}%
}                     % Do not remove
%
% \offprints{}          % Insert a name or remove this line
%
\institute{Department of Mathematics,\\ 
              Faculty of Science,\\ 
              The Maharaja Sayajirao University of Baroda,\\
              Vadodara  390 002, Gujarat, India.\\
              votmsu@gmail.com \and Department of Mathematics and Computer Science,\\ 
              Pandit Deendayal Petroleum University,\\
              Gandhinagar 382 007, Gujarat, India.\\
              dishantpandya777@gmail.com}
\date{Received: date / Revised version: date}
\abstract{
New exact solutions of Einstein's field equations (EFEs) by assuming linear equation of state, $ p_r = \alpha (\rho - \rho_R) $ where $ p_r $ is the radial pressure and $ \rho_R $ is the surface density, are obtained on the background of a paraboloidal spacetime. By assuming estimated mass and radius of strange star candidate 4U 1820-30, various physical and energy conditions are used for estimating the range of parameter $ \alpha $. The suitability of the model for describing pulsars like PSR J1903+327, Vela X-1, Her X-1 and SAX J1808.4-3658 has been explored and respective ranges of $ \alpha $, for which all physical and energy conditions are satisfied throughout the distribution, are obtained. 
 % end of PACS codes
} %end of abstract
\maketitle
\section{Introduction}
\label{intro}
Mathematical models of compact superdense stars such as pulsars and quark stars compatible with observational data has received wide attention in the recent past. A large number of research articles have emerged making different assumptions in the physical content as well as spacetime metrics. \citep{Murad15,Murad13a,Murad13b,Murad13c,Murad13d,Murad14a,Maurya11a,Maurya11b,Maurya11c,Maurya15a,Maurya16a,Maurya16b,Maurya16c,Maurya16d,Sharma13,Pandya14,Thomas15uncharged,Thomas15charged,Ratanpal15uncharged,Ratanpal15charged}. \\
Theoretical study of relativistic stars by \cite{Ruderman72} and \cite{Canuto74} have shown that when the matter distributions have density in the nuclear regime, the pressure distribution in the star may not be isotropic. Diverse reasons for the appearance of anisotropy have been extensively discussed by \cite{Bowers74}. Since then a number of research articles have been appeared in literature incorporating anisotropy in pressure. \citep{Maharaj89,Gokhroo94,Patel95,Tikekar98,Tikekar99,Tikekar05,Thomas05,Thomas07,Dev02,Dev03,Dev04}. \\
For constructing realistic relativistic models, the Einstein's field equations are to be solved by supplementing an equation of state (EOS) for the matter content. In many works recently appeared in literature, researchers used general barotropic equation of state in which the density and pressure related in linear, quadratic or polytropic form. In the construction of relativistic models compatible with observational data \cite{Sharma07} used linear equation of state. \cite{Ngubelanga15} also used linear equation of state in isotropic coordinates for physically viable relativistic models of compact stars. \cite{Feroze11} and \cite{Maharaj12} have used quadratic EOS for obtaining solution of anisotropic distributions.\cite{Thirukkanesh12,Maharaj13b} have used polytropic EOS for generating solutions for relativistic stars. \\
When the star consists of matter distribution beyond nuclear regime, the corresponding solution to EFEs is to be examined carefully. The general condition such as regularity, energy and causality conditions satisfied by the solution in the relativistic set up have been stipulated by \cite{Knutsen87,Murad15}. Once an EOS has been specified, TOV equation can be integrated from centre to boundary, where the pressure drops to zero, to determine the mass and radius of the star. For superdense stars like pulsars in the category of strange stars, linear equation of state is the most appropriate EOS for its matter distribution \cite{Sharma07}. They have studied relativistic stars with linear equation of state by taking the coefficient of $ dr^2 $ in the spacetime metric a specific form $ e^\mu  = \frac{1 + ar^2}{1+ (a-b)r^2}. $ Different solutions have been generated in this set up for different choices of arbitrary constants $ a $ and $ b $. Recently \cite{Ngubelanga15} obtained solutions for charged anisotropic distributions in isotropic coordinates with linear equation of state and compared their model with a number of known pulsars. \\
We have organized the content of the paper according to the following scheme. In section \ref{sec:2}, we have described the paraboloidal spacetime and the field equations assuming anisotropic matter distribution. The solution of field equations, assuming linear equation of state $ p = p(\rho) $, is obtained in section \ref{sec:3}. The three parameters of the model are $ A,L $ and $ \alpha $. The constant of integration $ A $ and the geometric parameter $ L $ of the spacetime, are obtained by matching the interior spacetime metric with Schwarzschild exterior metric across the boundary $ r = R $ in section \ref{sec:4}. In section \ref{sec:5}, we have obtained the bounds for the parameter $ \alpha $, appearing in the equation of state, by using the physical acceptability conditions at the centre $ r = 0 $ and on the boundary $ r = R $ by choosing $ M = 1.58 M_\odot $ and $ R = 9.1 km $, the mass and radius of the pulsar 4U 1820-30. We have verified the physical acceptability conditions using graphical methods in section \ref{sec:6} for the range of $ \alpha $ obtained from section \ref{sec:5}. The suitability of the model for describing pulsars like PSR J1903+327, Vela X-1, Her X-1, SAX J1808.4-3658 is examined and the corresponding ranges of $ \alpha $ are obtained in the last section. \\
\section{The Spacetime Metric}
\label{sec:2}
A three-paraboloid immersed in a four dimensional Euclidean space has the 
Cartesian equation
\begin{equation}
x^2+y^2+z^2 = 2\omega R
 \label{three-pseudo spheroid equation}
\end{equation}
The $ \omega = constant $ sections are spheres while the sections $ x = constant $, $ y 
= constant, $ and $ z = constant,$ respectively, give 3-paraboloids. \\
\noindent On taking the parametrization
\begin{eqnarray}\label{parametrization}
\nonumber x & = & r sin\theta cos\phi, \\ \nonumber
y & = & r sin\theta sin\phi, \\ \nonumber
z & = & r cos\theta, \\
\omega & = & \frac{r^2}{2L},
\end{eqnarray}
\noindent the Euclidean metric
\begin{equation}
d\sigma^2 = dx^2 + dy^2 + dz^2 + d\omega^2
\label{sigma}
\end{equation}
\noindent takes the form
\begin{equation}
d\sigma^2 = \left({1 + \frac{r^2}{L^2}}\right)dr^2 + r^2 d\theta^2 + r^2 sin^2 \theta d\phi^2
\label{sigma in spherical coordinates}
\end{equation}
where $ L $ is a constant. This metric has been extensively studied by \cite{Jotania06}. \\
We shall take the interior spacetime metric for the anisotropic fluid distribution as
\begin{equation}
ds^2 = e^{\nu (r)} dt^2 - e^{\lambda (r)} dr^2 - r^2 d\theta^2 - r^2 sin^2 \theta d\phi^2 ,
\label{interior_metric}
\end{equation}
where  
\begin{equation}
e^{\lambda (r)} = 1 + \frac{r^2}{L^2}.
\label{eraisedtolambda}
\end{equation}

The constant $ \frac{1}{L^2} $ can be identified with the constant $ C $ of 
Finch and Skea spacetime metric \cite{Finch89}. We can also obtain the right hand side of equation (\ref{eraisedtolambda}) by taking $ a = b = 
\frac{1}{L^2} $ in the expression $ e^\mu  = \frac{1 + ar^2}{1+ (a-b)r^2} $ of the stellar model given by \cite{Sharma07}. It is to be noted that the metric function is well-behaved for $ a = b $, whereas the coefficient of $ dt^2 $, viz., $ e^{\gamma} $, obtained as a solution of Einstein's field equations is singular for $ a = b $. So the solution given by \cite{Sharma07} excludes the possibility $ a = b $ in their solution. \\

Following \cite{Maharaj89}, we write the energy-momentum tensor for anisotropic fluid distribution as 
\begin{equation}
T_{ij} = \left(\rho + p \right) u_i u_j - p g_{ij} + \sqrt{3}S \left[C_iC_j - 
\frac{1}{3}(u_iu_j - g_{ij})\right],
\label{b}
\end{equation}

where $\rho,p$ and $u_i$, respectively, denote the energy density, isotropic 
pressure and $4-$velocity of the fluid. $ S(r) $ denotes the magnitude of 
anisotropic stress and $ C^i = (0,e^{-\frac{\lambda}{2}},0,0). $ \\

The surviving components of energy-momentum tensor are:
\begin{equation}
T_0^0 = \rho, ~~~ T_1^1 = - \left(p + \frac{2S}{\sqrt{3}}\right), ~~~ T_2^2 = 
T_3^3  =  -\left(p - \frac{S}{\sqrt{3}}\right).
\label{d}
\end{equation}

The radial and tangential pressures are now given by 
\begin{eqnarray}
p_r &=& -T_1^1 = \left(p + \frac{2S}{\sqrt{3}}\right),\label{e}\\
p_\perp &=& -T_2^2 = \left(p - \frac{S}{\sqrt{3}}\right).\label{f}
\end{eqnarray}
so that
\begin{equation}
S = \frac{p_r - p_\perp}{\sqrt{3}}.
\label{g}
\end{equation}

The Einstein's field equations constitute the following set of three non-linear 
differential equations in terms of potentials $ \lambda $ and $ \nu $.
\begin{eqnarray}
8\pi \rho &=& \frac{1 - e^{-\lambda}}{r^2} + \frac{e^{-\lambda} 
{\lambda}^\prime}{r},\label{i}\\
8\pi p_r &=& \frac{e^{-\lambda} - 1}{r^2} + 
\frac{e^{-\lambda}{\nu}^\prime}{r},\label{j}\\
8\pi p_\perp &=& e^{-\lambda} \left[\frac{\nu^{\prime\prime}}{2} + 
\frac{{\nu^\prime}^2}{4} - \frac{\nu^\prime \lambda^\prime}{4} + 
\frac{\nu^\prime - \lambda^\prime}{2r}\right],\label{k}
\end{eqnarray}

If we define
\begin{equation}
m(r) = 4\pi \int\limits_{0}^{r} x^2 \rho(x) dx,
\label{o}
\end{equation}

then the system of equations (\ref{i})--(\ref{k}) can be equivalently written 
as 
\begin{eqnarray}
e^{-\lambda} &=& 1 - \frac{2m}{r},\label{l}\\
\left(1 - \frac{2m}{r}\right)\nu^{\prime} &=& 8\pi p_r r + 
\frac{2m}{r^2},\label{m}\\
-\frac{4}{r} (8\pi \sqrt{3}S) &=& (8\pi\rho + 8\pi p_r)\nu^{\prime} + 2 (8\pi 
p_r^{\prime}).\label{n}
\end{eqnarray}

Using (\ref{eraisedtolambda}) in (\ref{i}), we get
\begin{equation}
8 \pi \rho = \frac{1}{L^2} \frac{\left(3 + \frac{r^2}{L^2}\right)}{\left(1 + 
\frac{r^2}{L^2}\right)^2},
\label{p}
\end{equation}
as the matter density and from equation (\ref{o}), we get
\begin{equation}
m(r) = \frac{1}{2L^2} \frac{r^3}{1 + \frac{r^2}{L^2}}.
\label{q}
\end{equation}
as the mass of the distribution inside the radius $ r $. \\

The metric potential $ \nu $ can be obtained from equation (\ref{m}) once we 
know the expression for $ p_r $. For this, we define an equation of state $ p_r 
= p_r (\rho). $ If we consider models of pulsar to be strange stars the most 
appropriate equation of state is linear equation of state considered by \cite{Dey98,Gondek00} and \cite{Zdunik00}. \\
\section{Linear Equation of State}
\label{sec:3}
We shall take 
\begin{equation}
 p_r = \alpha \rho + \beta
 \label{lineareos1}
\end{equation}
where $ \alpha $ and $ \beta $ are constants. The radius $ R $ of the star with this 
pressure distribution is obtained by using the condition $ p_r (r=R) = 0 $. 
This gives $ \beta = - \alpha \rho_R $, where $ \rho_R = \rho (r = R) $. \\
Therefore, equation (\ref{lineareos1}) takes the form 
\begin{equation}
 p_r = \alpha (\rho - \rho_R).
 \label{lineareos2}
\end{equation}

Using (\ref{lineareos2}) in equation (\ref{m}), we get 
\begin{equation}
 \nu'= \left\lbrace 
\alpha\left[\frac{3+\frac{r^2}{L^2}}{1+\frac{r^2}{L^2}} - 
\frac{3+\frac{R^2}{L^2}}{\left(1+\frac{R^2}{L^2}\right)^2}{\left(1+\frac{r^2}{
L^2 }\right)} \right] + 1 \right\rbrace\frac{r}{L^2}
\label{nudash}
\end{equation}

and consequently
\begin{eqnarray}
 \nonumber e^{\nu} = A \left(1 + \frac{r^2}{L^2}\right)^{\alpha} 
exp\left[\frac{(\alpha+1)}{2} \frac{r^2}{L^2} - \frac{\alpha}{2} \left(3 
+ \frac{R^2}{L^2}\right) \times \right. \\
\left. \left(1+\frac{R^2}{L^2}\right)^{-2} \left(1 + \frac{1}{2}\frac{r^2}{L^2}\right) \frac{r^2}{L^2}\right],~~~
\label{eraisedtonu}
\end{eqnarray}
where $ A $ is a constant of integration. This is a new exact solution which 
can not be obtained as a special case by putting $ a = b = \frac{1}{L^2} $ in 
the solution given by \cite{Sharma07}, because in their case 
when $ a = b $, the coefficient of $ dt^2 $ given by $ e^{\gamma} $ becomes 
zero.

The gradient of radial pressure is given by 
\begin{equation}
 8 \pi \frac{dp_r}{dr} = \alpha (8 \pi \frac{d\rho}{dr}) = - \alpha \frac{5 
+ \frac{r^2}{L^2}}{L^2 \left(1+\frac{r^2}{L^2}\right)^3}\frac{2r}{L^2} < 0.
\label{dprbydr}
\end{equation}

Hence the density $ \rho $ and pressure $ p_r $ are decreasing functions of $ r 
$. The anisotropy $ S $ has the expression
\begin{eqnarray}
\nonumber 8\pi \sqrt{3}S = -\frac{r^2}{L^2}\left\lbrace 
\frac{\alpha}{L^2}\frac{5+\frac{r^2}{L^2}}{\left(1+\frac{r^2}{L^2}\right)^3} \right. \\ \nonumber
\left. +\left[\frac{1+\alpha}{L^2}\frac{3+\frac{r^2}{L^2}}{\left(1+\frac{r^2}{L^2}
\right)^2}-\frac{\alpha}{L^2}\frac{3+\frac{R^2}{L^2}}{\left(1+\frac{R^2}{L^2}
\right)^2}\right] \times  \right. \\ 
 \left. \left[\frac{\alpha}{4}\frac{3+\frac{r^2}{L^2}}{\left(1+\frac{r^2}{L^2}
\right)}-\frac{\alpha}{4}\frac{3+\frac{R^2}{L^2}}{\left(1+\frac{R^2}{L^2}
\right)^2}\left(1+\frac{r^2}{L^2}
\right)+\frac{1}{4}\right]\right\rbrace
 \label{anisotropy}
\end{eqnarray}

It can be noticed from equation (\ref{anisotropy}) that the anisotropy vanishes 
at the centre. The tangential pressure $ p_\perp $ can be obtained using 
\begin{equation}
 8 \pi p_\perp = 8\pi p_r - 8\pi \sqrt{3}S
 \label{pperp}
\end{equation}

The expressions for $ \frac{dp_r}{d\rho} $ and $ \frac{dp_\perp}{d\rho} $ take 
the following form:
\begin{equation}
 \frac{dp_r}{d\rho} = \alpha
 \label{dprbydrho}
\end{equation}
% \begin{eqnarray}
% \nonumber \frac{dp_\perp}{d\rho} = \frac{1}{4} \frac{1}{L^{14} \left(1+\frac{r^2}{L^2}\right) \left(5+\frac{r^2}{L^2}\right) \left(1+\frac{R^2}{L^2}\right)^4} \\
%  \left(-L^{14}\left(3-\frac{r^2}{L^2}\right) \left(\frac{r^2}{L^2}+1\right) \left(\frac{R^2}{L^2}+1\right)^4+ \right. \\
% \left. 2 L^4 \left(\frac{R^2}{L^2}+1\right)^2 \left(L^{10} \left(\frac{3 r^8}{L^8}+\frac{12 r^6}{L^6}+\frac{23 r^4}{L^4}+\frac{16 r^2}{L^2}+20\right) \right. \left. -L^8 R^2 \left(\frac{r^8}{L^8}+\frac{4 r^6}{L^6}+\frac{11 r^4}{L^4}+\frac{2 r^2}{L^2}+30\right)\\
%  \label{dpperpbydrho}
% \end{eqnarray}

\begin{eqnarray}
\nonumber \frac{dp_\perp}{d\rho} = \frac{1}{4} \frac{1}{L^{14} \left(1+\frac{r^2}{L^2}\right) \left(5+\frac{r^2}{L^2}\right) \left(1+\frac{R^2}{L^2}\right)^4} \\ \nonumber 
-L^{14} \left(1+\frac{r^2}{L^2}\right) \left(3-\frac{r^2}{L^2}\right) \left(1+\frac{R^2}{L^2}\right)^4 + \\ \nonumber
  \alpha^2 L^{10} R^2 \left(1-\frac{r^2}{R^2}\right) \left(\frac{r^2 R^2}{L^4}+\frac{3 \left(r^2+R^2\right)}{L^2}+5\right)  \\ \nonumber
 \left(\frac{2 r^6 R^2}{L^6}-\frac{6 r^6+5 r^4 R^2+3 r^2 R^4}{L^4}-\frac{15 r^4+10 r^2 R^2-3 R^4}{L^2}+ \right. \\ \nonumber
 \left. 5 R^2 \left(\frac{3 r^2}{R^2}-1\right)\right)  + 2 \alpha  L^4 \left(1+\frac{R^2}{L^2}\right)^2  \\ \nonumber 
 \left(-2 L^6 R^4 \left(\frac{r^4}{L^4}-\frac{r^2}{L^2}+7\right)-L^8 R^2  \right. \\ \nonumber
 \left. \left(\frac{r^8}{L^8}+\frac{4 r^6}{L^6}+\frac{11 r^4}{L^4}+\frac{2 r^2}{L^2}+30\right)+ \right. \\ \nonumber
 \left. L^{10} \left(\frac{3 r^8}{L^8}+\frac{12 r^6}{L^6}+\frac{23 r^4}{L^4}+\frac{16 r^2}{L^2}+20\right)\right) \\
 \label{dpperpbydrho}
\end{eqnarray}

The unknown parameters in our model with linear equation of state are $ A, 
\alpha $ and $ L $.

\section{Matching Condition}
\label{sec:4}

The matching of first fundamental form across the boundary $ r = R $ guarantees the continuity of the metric coefficients across $ r = R $. On matching the interior spacetime metric (\ref{interior_metric}) with the Schwarzschild exterior metric 
\begin{equation}
 ds^2 = \left(1 - \frac{2M}{r}\right) dt^2 - \left(1 - 
\frac{2M}{r}\right)^{-1}dr^2 - r^2 d\theta^2 - r^2 sin^2 \theta d\phi^2,
 \label{SchwarzschildExtMetric}
\end{equation}
we get 
\begin{equation}
 1 - \frac{2M}{R} = \left(1 + \frac{R^2}{L^2}\right)^{-1}
 \label{OneminustwoMbyR_1}
\end{equation}
and
\begin{equation}
 1 - \frac{2M}{R} = A \left(1 + \frac{R^2}{L^2}\right)^{\alpha}e^{\frac{1}{2}\frac{R^2}{L^2}(\alpha + 1) - \frac{3 + \frac{R^2}{L^2}}{\left(1 + \frac{R^2}{L^2}\right)^2} \frac{\alpha}{2}\frac{R^2}{L^2}\left(1 + \frac{1}{2} \frac{R^2}{L^2}\right)}.
 \label{OneminustwoMbyR_2}
\end{equation}
Equations (\ref{OneminustwoMbyR_1}) and (\ref{OneminustwoMbyR_2}) determine the geometric parameter $ L $ and the constant of integration $ A $ as
\begin{equation}
L = \sqrt{\frac{R^3}{2M}\left(1-\frac{2M}{R}\right)},
\label{L} 
\end{equation}
\begin{eqnarray}
\nonumber A = \frac{1}{\left(1+\frac{R^2}{L^2}\right)^{\alpha +1}} \times \\
exp \left\lbrace 
-\frac{1}{2}\frac{R^2}{L^2} (\alpha + 1) + 
\frac{3+\frac{R^2}{L^2}}{\left(1+\frac{R^2}{L^2}\right)^2 
}\frac{\alpha}{2}\frac{R^2}{L^2}
\left(1+\frac{1}{2}\frac{R^2}{L^2}\right) \right\rbrace.~~~
 \label{A}
\end{eqnarray}
The second condition imposed on the boundary is that $ \frac{\partial g_{tt}}{\partial r} $ of the interior spacetime metric (\ref{interior_metric}) should match continuously with that of exterior spacetime metric (\ref{SchwarzschildExtMetric}) across $ r = R $. \citep{Misner64,Maurya17}. This guarantees the continuity of radial pressure across the boundary $ r = R $. It is found that $ \frac{\partial g_{tt}}{\partial r} $ at $ r = R $ takes the value $ \frac{R}{L^2 \left(1 + \frac{R^2}{L^2}\right)} $ for both metrics (\ref{interior_metric}) and (\ref{SchwarzschildExtMetric}) indicating the continuity of the derivative of metric coefficients and in turn the continuity of radial pressure.
% At the boundary $ r = R $, the interior spacetime metric (\ref{interior_metric}) 
% should continuously match with the Schwarzschild exterior metric
% \begin{equation}
%  ds^2 = \left(1 - \frac{2M}{r}\right) dt^2 - \left(1 - 
% \frac{2M}{r}\right)^{-1}dr^2 - r^2 d\theta^2 - r^2 sin^2 \theta d\phi^2.
%  \label{SchwarzschildExtMetric}
% \end{equation}
% 
% The matching conditions determine the parameters $ L $ and $ A $ as 
% \begin{equation}
% L = \sqrt{\frac{R^3}{2M}\left(1-\frac{2M}{R}\right)},
% \label{L} 
% \end{equation}
% \begin{eqnarray}
% \nonumber A = \frac{1}{\left(1+\frac{R^2}{L^2}\right)^{\alpha +1}} \times \\
% exp \left\lbrace 
% -\frac{1}{2}\frac{R^2}{L^2} (\alpha + 1) + 
% \frac{3+\frac{R^2}{L^2}}{\left(1+\frac{R^2}{L^2}\right)^2 
% }\frac{\alpha}{2}\frac{R^2}{L^2}
% \left(1+\frac{1}{2}\frac{R^2}{L^2}\right) \right\rbrace~~~
%  \label{A}
% \end{eqnarray}
In order to validate our model for known star, we consider $ 4U 1820-30 $ whose mass and radius are respectively $ 1.58 M_\odot $ and $ 9.1 km $ \cite{Gangopadhyay13}. Using these values, we obtain the geometric parameter  $ L = 8.88 $ and consequently the integration constant $ A $ from equations (\ref{L}) and (\ref{A}), respectively. The range for the parameter $ \alpha $ can be determined using the physical acceptability conditions described in Section \ref{sec:5}.
\section{Physical Acceptability Conditions}
\label{sec:5}
Out of the $ 127 $ solutions examined by \cite{Delgaty98} only $ 16 $ 
solutions passed the elementary test for physical relevance out of which only 9 qualified the decreasing sound speed from centre to boundary. Hence it is pertinent to examine the following physical acceptability conditions \citep{Kuchowicz72,Buchdahl79,Murad15,Knutsen87} to validate the model. 
\renewcommand\theenumi{\textbf{\alph{enumi}}}
\renewcommand\theenumii{(\roman{enumii})}
\renewcommand\labelenumii{\theenumii}
\begin{enumerate}
 \item \textbf{Regularity Conditions:}
 \begin{enumerate}
\vspace{0.2cm}
\item The metric potentials $ e^\lambda > 0 $, $ e^{\nu} > 0 $ for $ 0 \leq r 
\leq R $. \\
        From equations (\ref{eraisedtolambda}) and (\ref{eraisedtonu}) it is 
clear that these conditions are indeed satisfied in the present model. \\
  \item $ \rho (r) \geq 0, ~~ p_r (r) \geq 0, ~~ p_\perp (r) \geq 0 $ for $ 0 
\leq r \leq R $. \\
        Equations (\ref{p}) and (\ref{lineareos2}), respectively, indicate that 
$ \rho \geq 0,~~p_r \geq 0 $ for $ 0 \leq r \leq R $. \\  
        From equations (\ref{anisotropy}) and (\ref{pperp}) it can be shown that the conditions $ p_{\perp} (r = 0) \geq 0 $ and $ p_{\perp} (r = R) \geq 0 $ impose a bound on $ \alpha $, viz., $ 0 \leq \alpha \leq 0.34311 $. \\
        \item $ p_r (r = R) = 0. $ \\
        Equation (\ref{lineareos2}), for radial pressure $ p_r $ clearly shows $ p_r (r = R) = \alpha (\rho_R - \rho_R) = 0 $, where $ R $ is the boundary radius of the star. \\
 \end{enumerate}
\item \textbf{Causality Conditions:}
\vspace{0.2cm}
\begin{enumerate}
  \item $ 0 \leq \frac{dp_r}{d\rho} \leq 1,~ 0 \leq 
  \frac{dp_\perp}{d\rho} \leq 1 $ for $ 0 \leq r \leq R $. \\
  From equation (\ref{lineareos2}), $ \frac{dp_r}{d\rho} = \alpha $ and consequently $ 0 \leq \frac{dp_r}{d\rho} \leq 1 $ implies $ 0 \leq \alpha \leq 1 $. \\
  \end{enumerate}
 \item \textbf{Energy Conditions:}
\vspace{0.2cm} 
 \begin{enumerate}
  \item $ \rho - p_r - 2p_\perp \geq 0 $ (strong energy condition), $ \rho \geq p_r $ and $ \rho \geq p_\perp $ (weak energy condition) for $ 0 \leq r \leq R $. \\ 
  $ \rho - p_r - 2p_\perp \geq 0 $ at $~r = 0 $ and $ r = R $ impose conditions on $\alpha,$ viz., $\alpha \leq 0.491062$ and $ -0.310332 \leq \alpha $). \\
 \end{enumerate}
 \vspace{0.2cm}
 \item \textbf{Monotone Decrease of Physical Parameters}
 \vspace{0.2cm}
 \begin{enumerate}
  \item $ \frac{d\rho}{dr} \leq 0,~\frac{dp_r}{dr} \leq 0,$ for $ 0 \leq r \leq R $. \\
  By equation (\ref{dprbydr}), the gradients of density and radial pressure are given by $ 8\pi \frac{dp_r}{dr} = \alpha (8\pi \frac{d\rho}{dr}) = - \alpha \frac{5+\frac{r^2}{L^2}}{L^2 \left(1+\frac{r^2}{L^2}\right)^3}\frac{2r}{L^2} \leq 0 $ for $ 0 \leq r \leq R $, \\
   indicating that the density and radial pressure are decreasing radially outward. \\

  \item $ \frac{d}{dr}\left(\frac{dp_r}{d\rho}\right) \leq 0,$ for $ 0 \leq r \leq R $.\\
  From equation (\ref{dprbydrho}), it is evident that
  \[\frac{d}{dr} \left(\frac{dp_r}{d\rho}\right) = 0 \] 
  indicating that the stipulated condition indeed holds.
  
  \item $ \frac{d}{dr}\left(\frac{p_r}{\rho}\right) \leq 0$ for $ 0 \leq r \leq R $. \\
       From equation(\ref{lineareos2}), we have \[\frac{d}{dr}\left(\frac{p_r}{\rho}\right) = \frac{d}{dr} \alpha \left(1 - \frac{\rho_R}{\rho}\right) = \alpha \frac{\rho_R}{\rho^2}\frac{d\rho}{dr}. \] Since $ \frac{d\rho}{dr} < 0, $ we have $ \frac{d}{dr}\left(\frac{p_r}{\rho}\right) < 0 $, indicating that $ \frac{p_r}{\rho} $ is a decreasing function of $ r $.\\
 \end{enumerate}
 \item \textbf{Pressure Anisotropy}
 \vspace{0.2cm}
  \begin{enumerate}
  \item $ S(r=0) = 0 $. \\
  It can be observed from equation (\ref{anisotropy}) that the pressure anisotropy $ S $ vanishes at the centre $ r = 0 $. \\
  \end{enumerate}
\vspace{0.2cm}
 \item \textbf{Mass-Radius Relation:}
 \begin{enumerate}
  \item  According to \cite{Buchdahl79}, the allowable mass radius ratio must satisfy the inequality $ \frac{M}{R} \leq \frac{4}{9} $. For the present model $ \frac{M}{R} = 0.256 < \frac{4}{9} $. \\
 \end{enumerate}
\vspace{0.2cm} 
 \item \textbf{Redshift}
 \begin{enumerate}
  \item $ z = e^{-\frac{\nu}{2}} - 1 $ must be decreasing and finite for $ 0 \leq r \leq R $. \\
  It is observed that $ \frac{dz}{dr}_{(r = 0)} = 0 $ and $ \frac{dz}{dr}_{(r = R)} = -0.082619 $. \\
 \end{enumerate}
 \vspace{0.2cm}
 \item \textbf{Stability Conditions:}
A model for which $ -1 \leq v_\perp^2 - v_r^2 \leq 0, $ where $ v_r $ and $ v_\perp $ denote radial and transverse sound speed, is potentially stable. This is equivalent to showing that $ 0 \leq \frac{dp_r}{d\rho} - \frac{dp_\perp}{d\rho} \leq 1. $ \citep{Abreu07,Maurya17}. Hence,
 \begin{enumerate}
%     $ 0 \leq \left(\frac{dp_r}{d\rho}\right) - \left(\frac{dp_\perp}{d\rho}\right) \leq 1 $ for $ 0 \leq r \leq R $. \\
   \item $ 0 \leq \left(\frac{dp_r}{d\rho}\right) - \left(\frac{dp_\perp}{d\rho}\right) \leq 1 $ at $ r = 0 $ and $ r = R $, respectively, give the bounds for $ \alpha $, viz., $ -1.04907 \leq \alpha \leq 0.280618 $ and $ -2.19921 \leq \alpha \leq 0.192717 $. \\
   To verify the above condition throughout the star we use graphical techniques, which we postpone to next section.
  \item The relativistic adiabatic index $ \Gamma = 
\frac{\rho+p_r}{p_r}\frac{dp_r}{d\rho} > \frac{4}{3}$. for $ 0 \leq r \leq R $. \\
$ \Gamma > \frac{4}{3} $ at $ r = 0 $ imposes a restriction on $ \alpha $ given by $ \alpha > -0.139852. $ and at the boundary it is automatically satisfied.
  \end{enumerate}
\end{enumerate}
Considering all the bounds on $ \alpha $ obtained from the physical acceptability conditions (a) - (h), the valid range of $ \alpha $ is obtained as $ 0.157156 \leq \alpha \leq 0.192717 $ in which all the conditions are satisfied without fail.
\begin{figure}
\resizebox{0.5\textwidth}{!}{
	\includegraphics{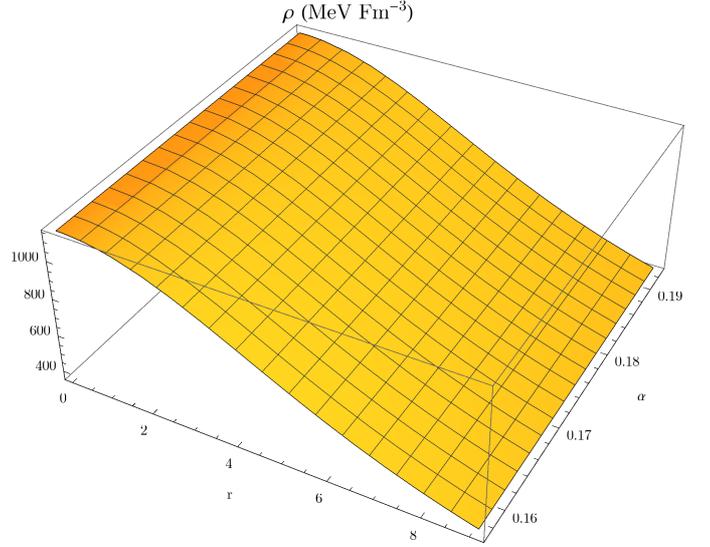}}
    \caption{Variation of a density $ \rho $ in MeV Fm$^{-3}$ with respect to a radial coordinate $ r $ for a star 4U 1820-30 within a range [0,9.1] kms and a constant $ \alpha $ in the range $ [0.157156,0.192717] $.}
    \label{fig:Density_figure}
\end{figure}
\begin{figure}
\resizebox{0.5\textwidth}{!}{
	\includegraphics{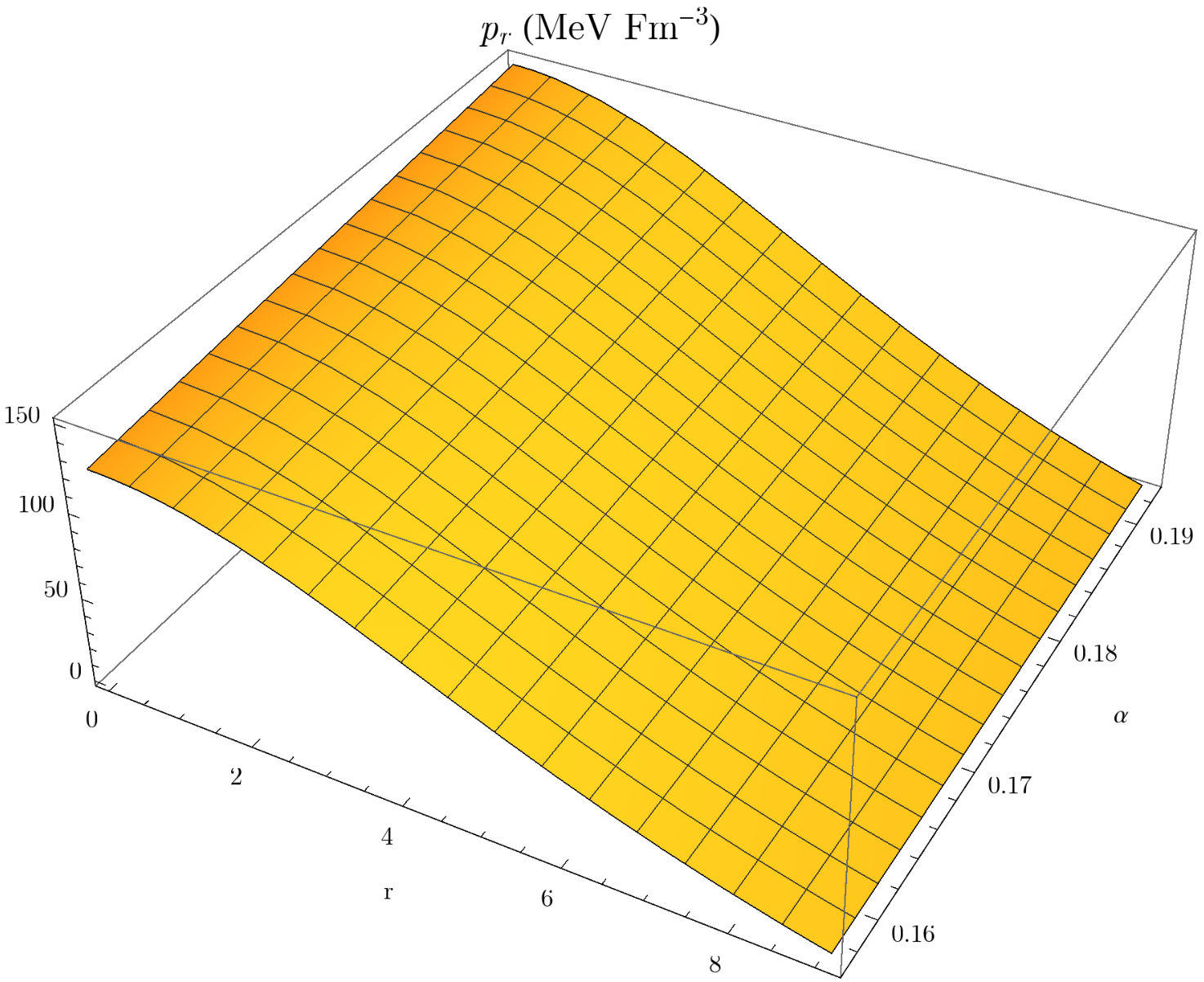}}
    \caption{Variation of a radial pressure $ p_r $ in MeV Fm$^{-3}$ with respect to a radial coordinate $ r $ for a star 4U 1820-30 within a range [0,9.1] kms and a constant $ \alpha $ in the range $ [0.157156,0.192717] $.}
    \label{fig:Radial_Pressure_figure}
\end{figure}
\begin{figure}
\resizebox{0.5\textwidth}{!}{
	\includegraphics{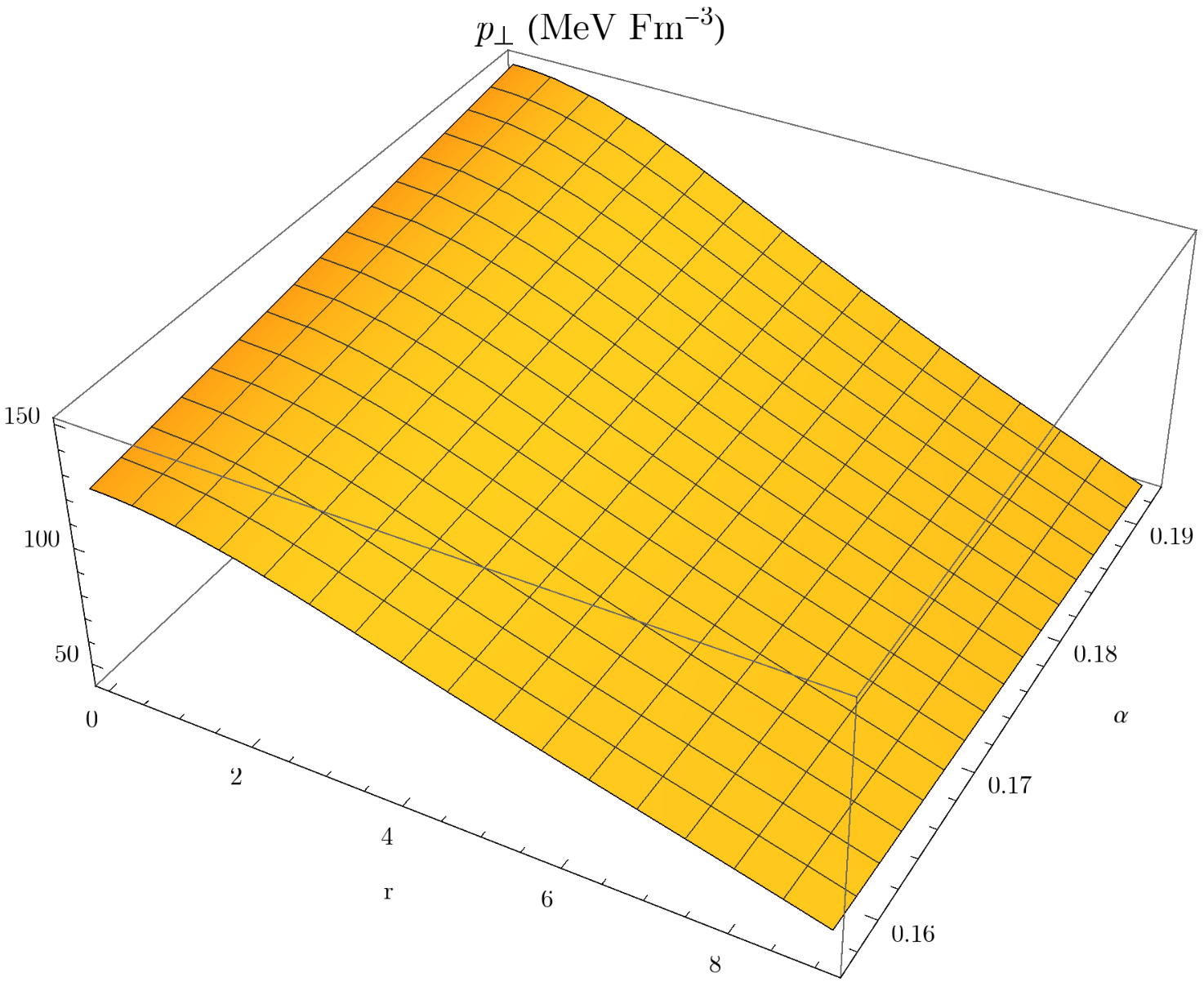}}
    \caption{Variation of a transverse pressure $ p_\perp $ in MeV Fm$^{-3}$ with respect to a radial coordinate $ r $ for a star 4U 1820-30 within a range [0,9.1] kms and a constant $ \alpha $ in the range $ [0.157156,0.192717] $.}
    \label{fig:Transverse_Pressure_figure}
\end{figure}
\begin{figure}
\resizebox{0.5\textwidth}{!}{
	\includegraphics{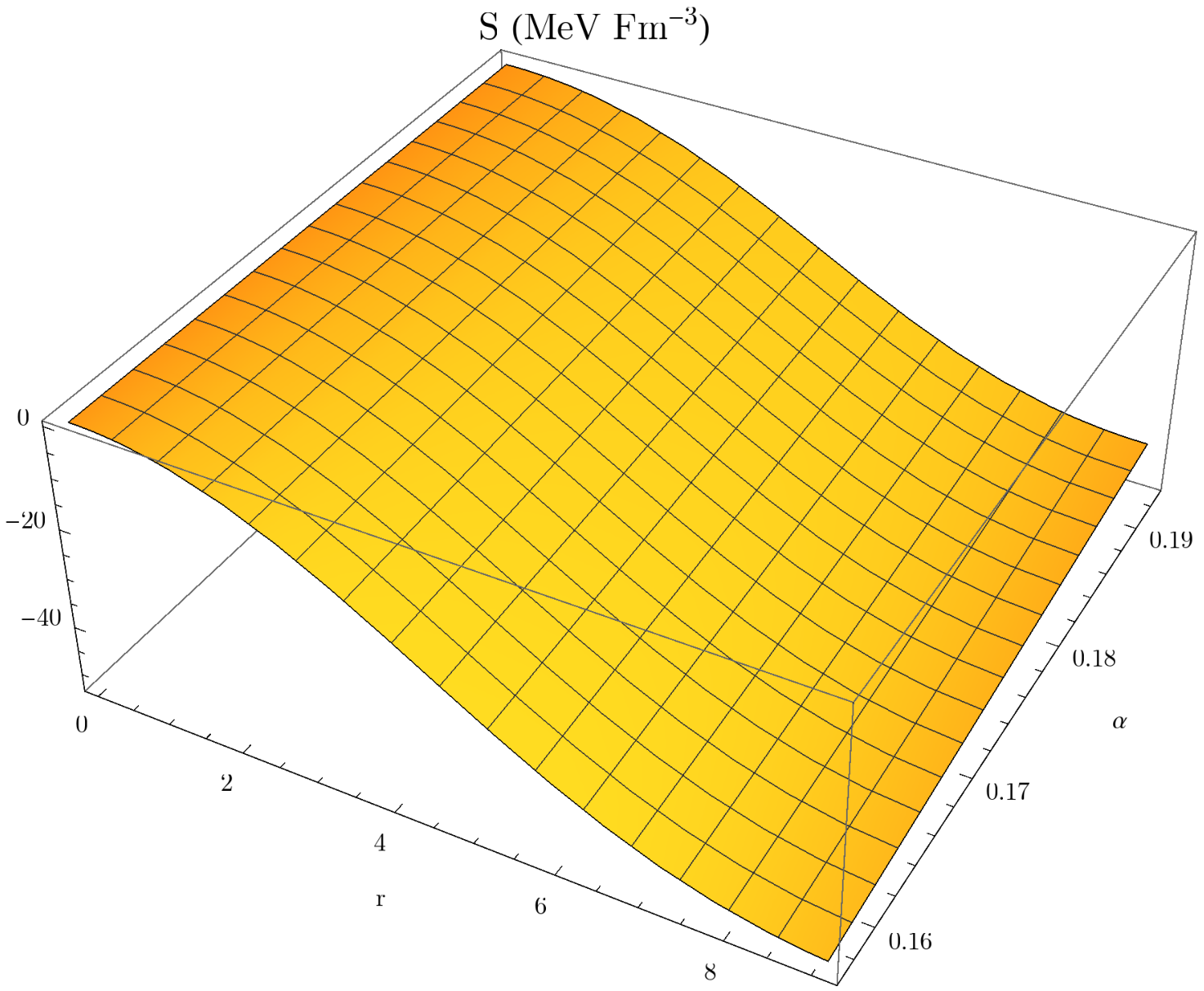}}
    \caption{Variation of an anisotropy $ S $ in MeV Fm$^{-3}$ with respect to a radial coordinate $ r $ for a star 4U 1820-30 within a range [0,9.1] kms and a constant $ \alpha $ in the range $ [0.157156,0.192717] $.}
    \label{fig:Anisotropy_figure}
\end{figure}
\begin{figure}
\resizebox{0.5\textwidth}{!}{
	\includegraphics{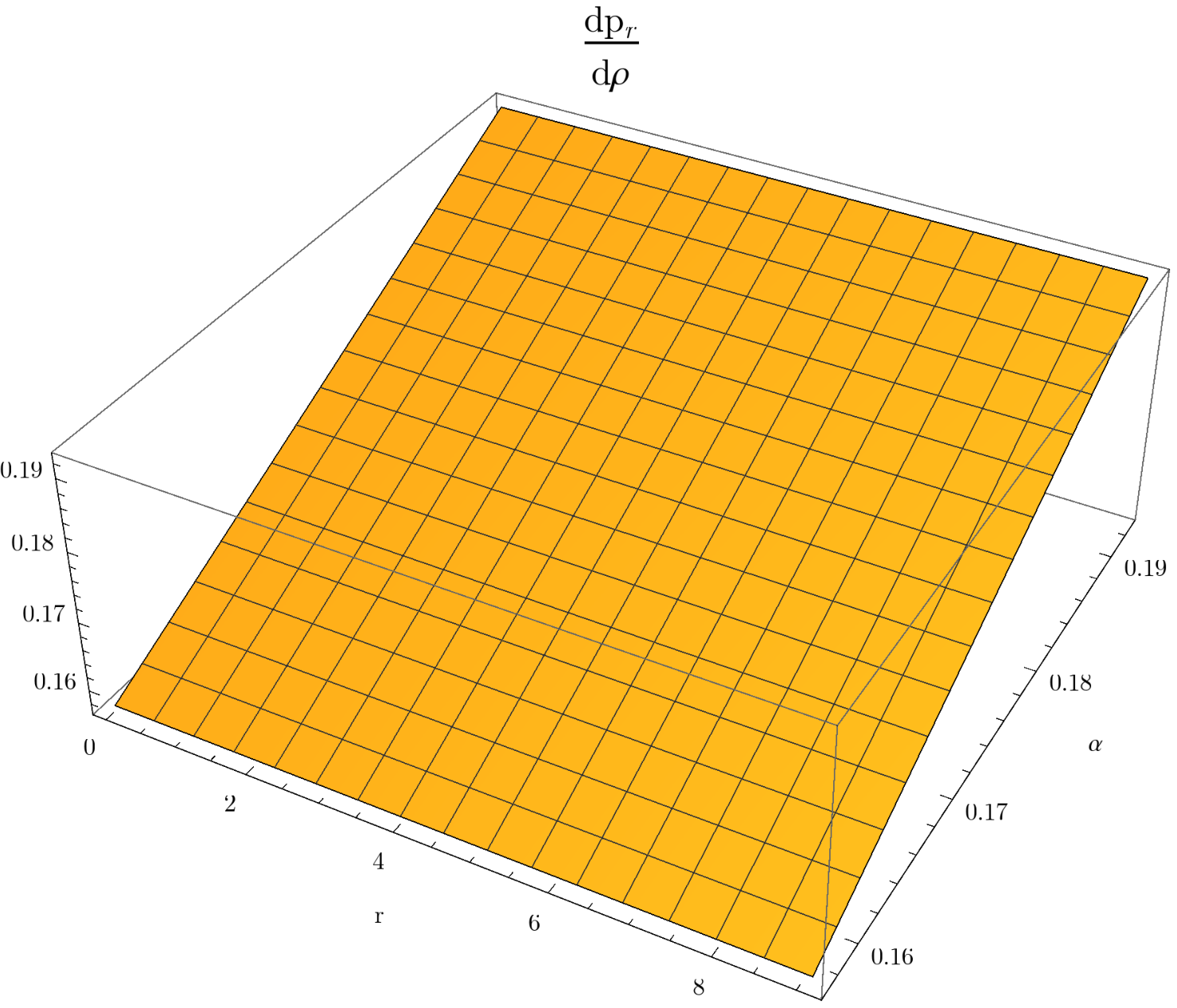}}
    \caption{Variation of a radial sound speed $ \frac{dp_r}{d\rho} $ with respect to a radial coordinate $ r $ for a star 4U 1820-30 within a range [0,9.1] kms and a constant $ \alpha $ in the range $ [0.157156,0.192717] $.}
    \label{fig:Radial_Sound_Speed_figure}
\end{figure}
\begin{figure}
\resizebox{0.5\textwidth}{!}{
	\includegraphics{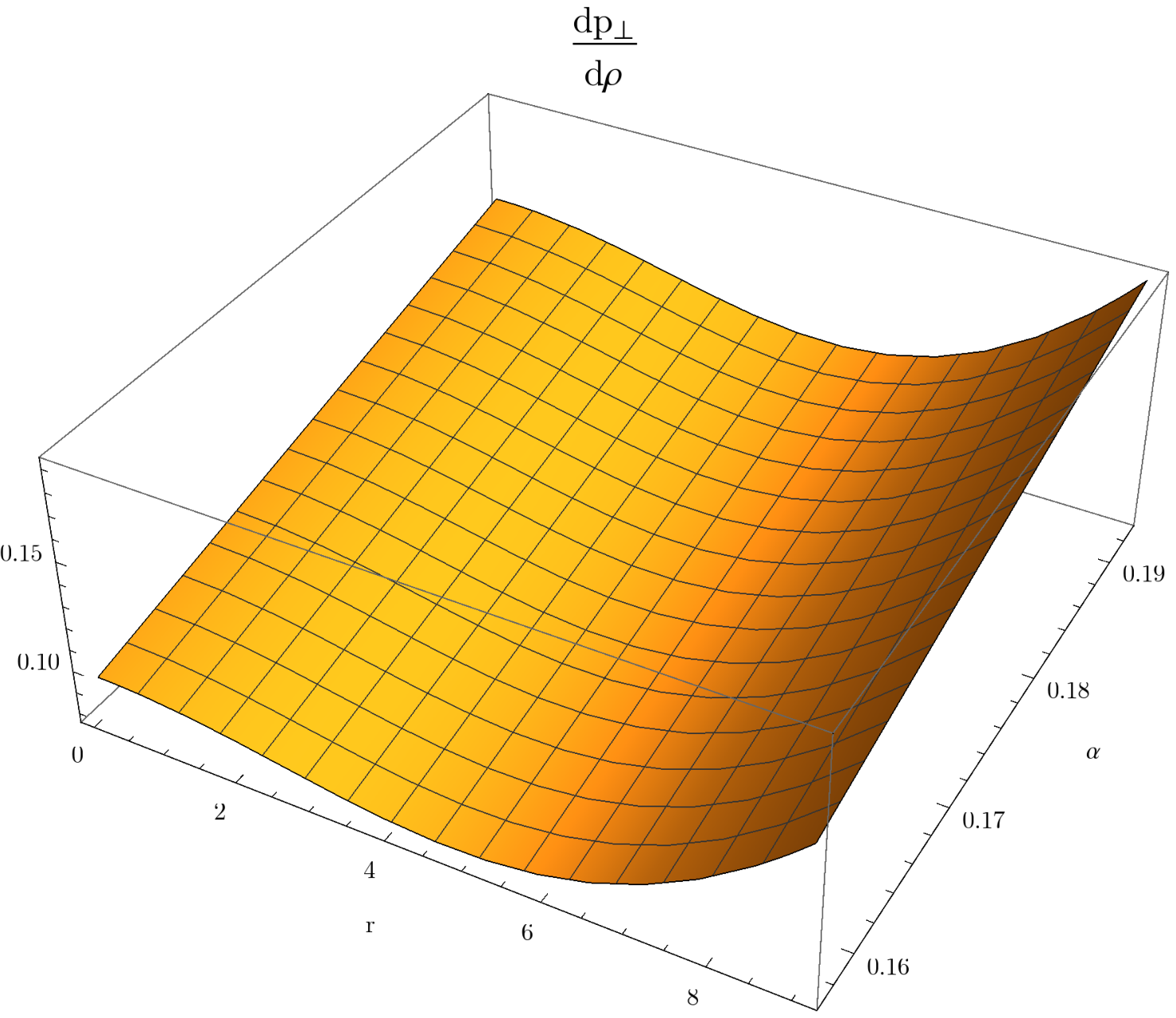}}
    \caption{Variation of a transverse sound speed $ \frac{dp_\perp}{d\rho} $ with respect to a radial coordinate $ r $ for a star 4U 1820-30 within a range [0,9.1] kms and a constant $ \alpha $ in the range $ [0.157156,0.192717] $.}
    \label{fig:Transverse_Sound_Speed_figure}
\end{figure}
\begin{figure}
\resizebox{0.5\textwidth}{!}{
	\includegraphics{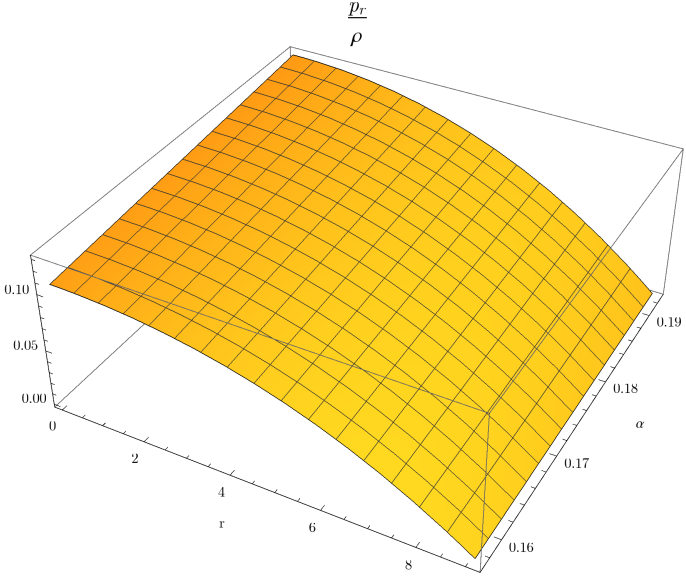}}
    \caption{Variation of a ratio $ \frac{p_r}{\rho} $ with respect to a radial coordinate $ r $ for a star 4U 1820-30 within a range [0,9.1] kms and a constant $ \alpha $ in the range $ [0.157156,0.192717] $.}
    \label{fig:pr_by_rho_figure}
\end{figure}
\begin{figure}
\resizebox{0.5\textwidth}{!}{
	\includegraphics{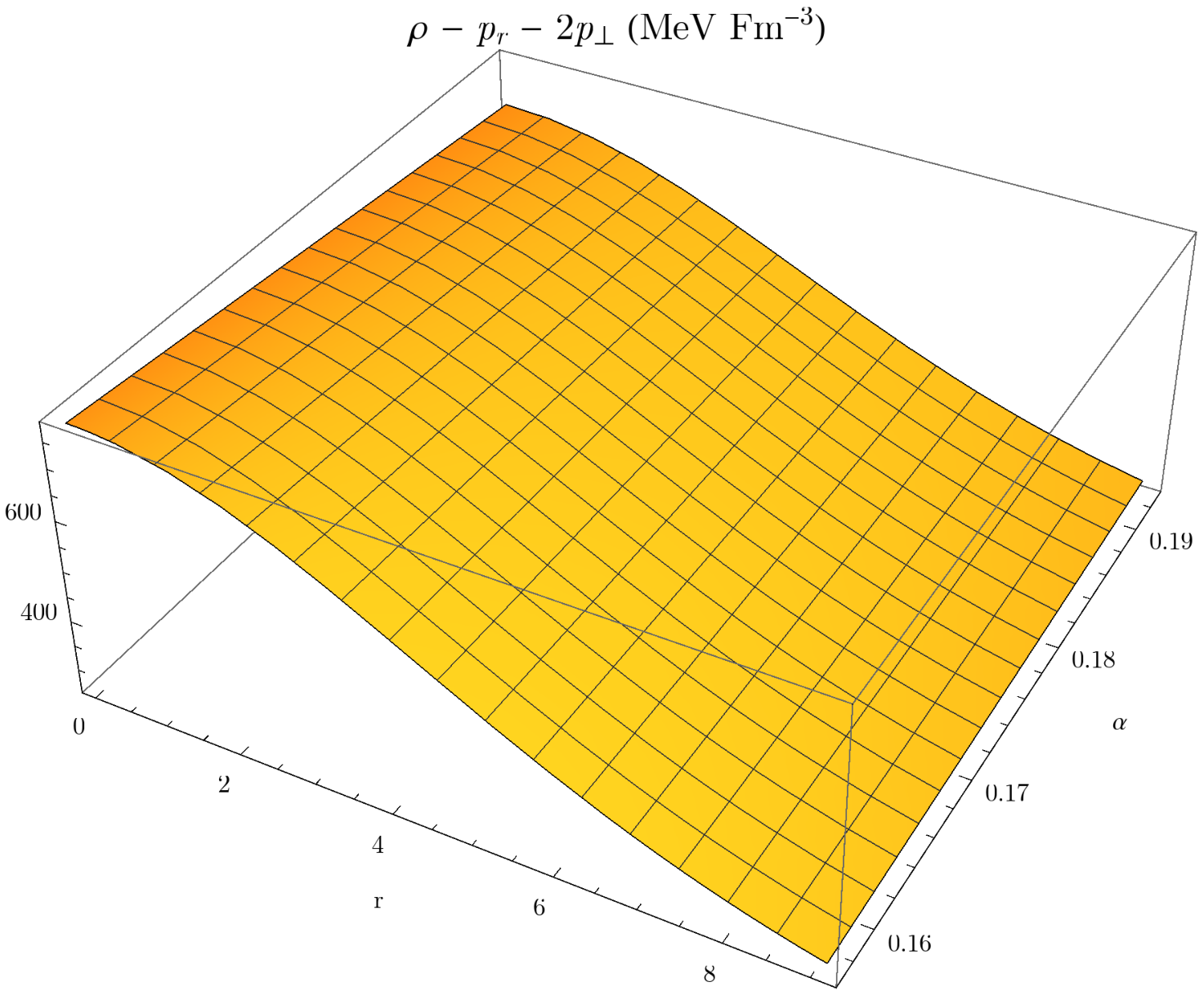}}
    \caption{Variation of a strong energy expression $ \rho - p_r - 2 p_\perp $ in MeV Fm$^{-3}$ with respect to a radial coordinate $ r $ for a star 4U 1820-30 within a range [0,9.1] kms and a constant $ \alpha $ in the range $ [0.157156,0.192717] $.}
    \label{fig:SEC_figure}
\end{figure}
\begin{figure}
\resizebox{0.5\textwidth}{!}{
	\includegraphics{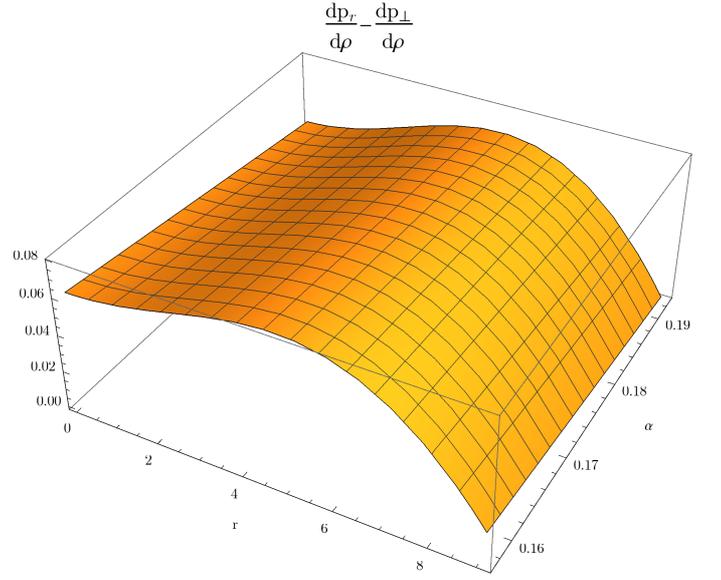}}
    \caption{Variation of a stability expression $ \left(\frac{dp_r}{d\rho} - \frac{dp_\perp}{d\rho} \right) $ with respect to a radial coordinate $ r $ for a star 4U 1820-30 within a range [0,9.1] kms and a constant $ \alpha $ in the range $ [0.157156,0.192717] $.}
    \label{fig:Stability_figure}
\end{figure}
\begin{figure}
\resizebox{0.5\textwidth}{!}{
	\includegraphics{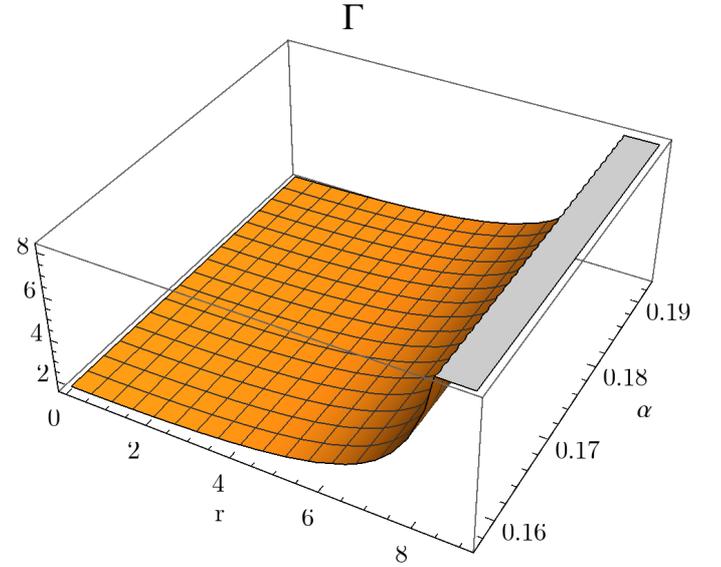}}
    \caption{Variation of an adiabatic index $ \Gamma $ with respect to a radial coordinate $ r $ for a star 4U 1820-30 within a range [0,9.1] kms and a constant $ \alpha $ in the range $ [0.157156,0.192717] $.}
    \label{fig:Adiabatic_Index_figure}
\end{figure}
\begin{figure}
\resizebox{0.5\textwidth}{!}{
	\includegraphics{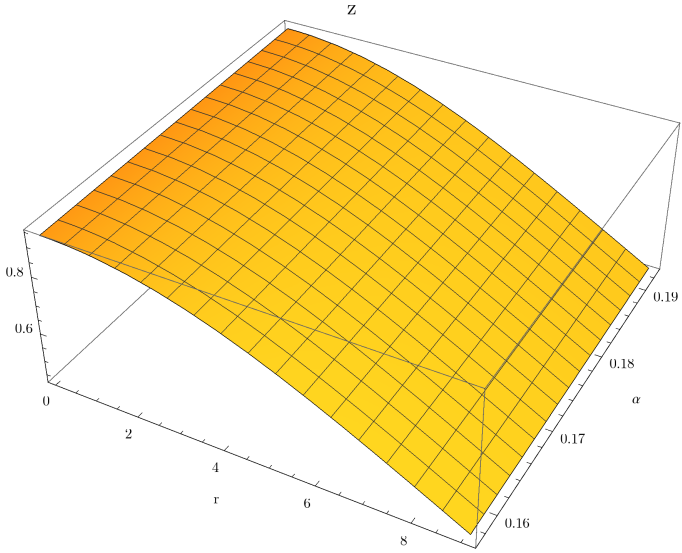}}
    \caption{Variation of a redshift $ z $ with respect to a radial coordinate $ r $ for a star 4U 1820-30 within a range [0,9.1] kms and a constant $ \alpha $ in the range $ [0.157156,0.192717] $.}
    \label{fig:Redshift_figure}
\end{figure}
\begin{figure}
\resizebox{0.5\textwidth}{!}{
	\includegraphics{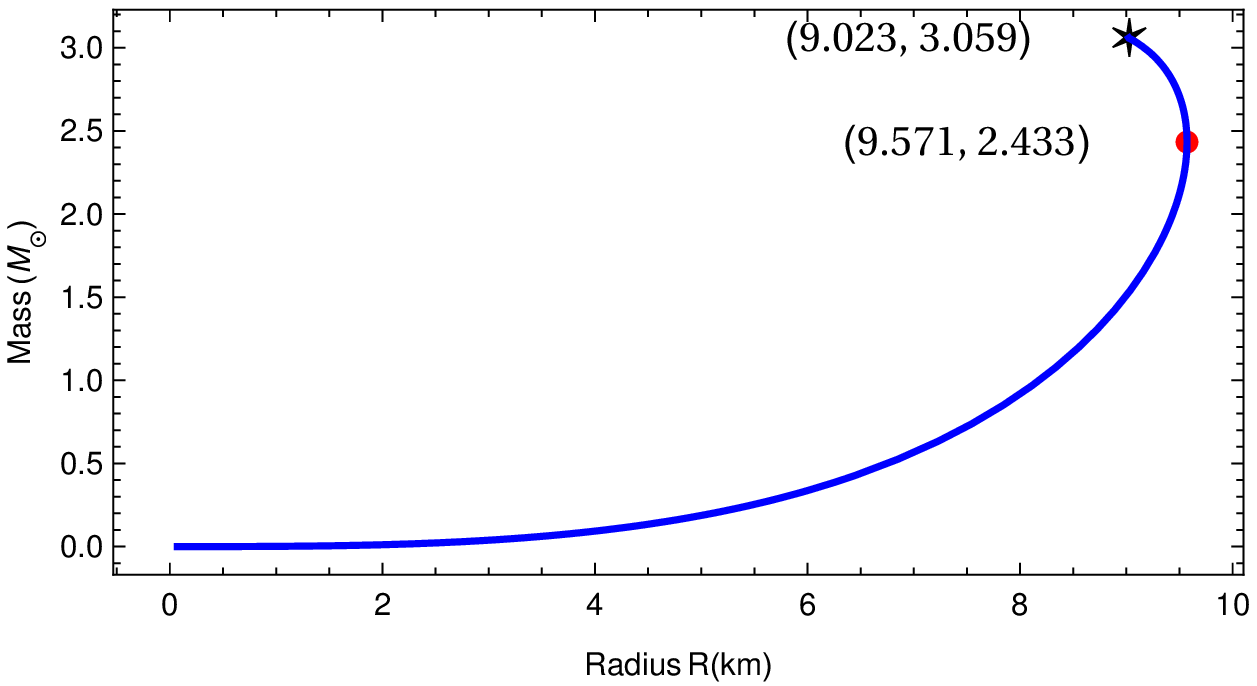}}
    \caption{Variation of a mass $ M $ within a range $ [0,3.059] M_\odot $ with respect to a radius $ R $ of a star 4U 1820-30 within a range [0,9.571] kms.}
    \label{fig:Mass-Radius Relationship}
\end{figure}
\section{Discussion}
\label{sec:6}
In order to examine the nature of various physical quantities throughout the distribution we shall adopt graphical method in the allowable range of $ \alpha $ viz., $ 0.157156 \leq \alpha \leq 0.192717 $. Figures \ref{fig:Density_figure}, \ref{fig:Radial_Pressure_figure} and \ref{fig:Transverse_Pressure_figure} clearly show that the density $ \rho $, radial pressure $ p_r $ and transverse pressure $ p_\perp $ are decreasing functions of radius $ r $. In many models found in literature the transverse pressure is not decreasing function of $ r $. Figure \ref{fig:Anisotropy_figure} indicates that the anisotropy is zero at the centre and decreasing throughout the distribution. \\
In Figure \ref{fig:Radial_Sound_Speed_figure}, we have shown the radial sound speed for $ \alpha $ in the range $ 0.157156 \leq \alpha \leq 0.192717, $ which is constant for any given $ \alpha $. Similarly, figure \ref{fig:Transverse_Sound_Speed_figure} represents how transverse sound speed varies with respect to radial variable $ r $ for $ 0.157156 \leq \alpha \leq 0.192717, $. Figure \ref{fig:pr_by_rho_figure} shows the decreasing behavior of the ratio $ \frac{p_r}{\rho} $ with the radial coordinate $ r $ and specified constant $ \alpha $ in the same given range.  
Figure \ref{fig:SEC_figure} depicts that the strong energy condition is satisfied throughout the distribution. In order that the relativistic model to represent a stable model, we must have $ 0 \leq \frac{dp_r}{d\rho} - \frac{dp_\perp}{d\rho} \leq 1 $ and the relativistic adiabatic index $ \Gamma > \frac{4}{3} $. Figures \ref{fig:Stability_figure} and \ref{fig:Adiabatic_Index_figure} clearly show that these conditions are satisfied throughout the distribution for $ 0.157156 \leq \alpha \leq 0.192717 $. \\ 
For the relativistic star, the redshift must be decreasing radially outward and finite throughout the distribution. Figure \ref{fig:Redshift_figure} shows that this is indeed the case throughout the star for the valid range of $ \alpha $. We have used the physical parameters of a known star 4U 1820-30 \citep{Gangopadhyay13} to validate the model. The redshift at the centre of the star 4U 1820-30 is given by $ z_0 = -1 + 1.86172 \sqrt {e^{0.47136\alpha}} $ and boundary redshift is given by $ z_R = 0.431839 $. It can be noticed that the central redshift is an increasing function of $ \alpha $. For $ 0.157156 \leq \alpha \leq 0.192717, $ the central redshift varies in the range $ [0.931968,0.948228]. $ We have calculated the central and boundary redshifts for PSR J1903+327, Vela X-1, Her X-1 and SAX J1808.4-3658 and displayed it in Table \ref{tab:1}. In particular, for the star Her X-1, the central redshift is $ z_0 = -1 + 1.34622 \sqrt {e^{0.143476\alpha}} $ and surface redshift $ z_R = 0.203473. $ For Her X-1, the range of $ \alpha $ is: $ 0.113357 \leq \alpha \leq 0.219847. $ Hence the central redshift $ z_0 $ is in the range $ \left[0.357209, 0.367617\right] $ which is in good agreement with \cite{Maurya15b}. For realistic anisotropic star models the surface redshift cannot exceed the values 3.842 or 5.211 when the tangential pressure satisfies the strong or dominant energy condition, respectively as suggested by \cite{Ivanov}; evidently the present model justifies this requirement immediately. \\
In Figure \ref{fig:Mass-Radius Relationship}, we have analyzed the mass-radius (M-R) relationship obtained from the model. The red dot in the figure represents the maximum radius of star and the star marker represents the maximum mass permitted by the model. From the figure it can be noticed that the maximum radius is 9.571 kms and the corresponding star mass is 2.433 $ M_\odot $, while the maximum permitted mass is 3.059 $ M_\odot $ and the corresponding radius is 9.023 kms.\\
The present model is in good agreement with the mass and size of the star 4U 1820-30 and satisfy all the physical acceptability conditions with $ \alpha $ in the range  $ 0.157156 \leq \alpha \leq 0.192717 $.
\section{Application of the Model to Other Stars}
\label{sec:7}
We have examined our model with stars like PSR J1903+327, Vela X-1, Her X-1, SAX J1808.4-3658 and found that the model is in good agreement with the mass and radius of these stars given by \cite{Gangopadhyay13}. The valid ranges of the parameter $ \alpha $ for which all the physical, regularity and energy conditions are satisfied, are displayed in Table \ref{tab:1}. \\
\onecolumn
\begin{table}[htbp]
\centering
\scriptsize
\caption{Bounds of $ \alpha $ for various physical conditions throughout the region $ 0 \leq r \leq R $.}
\label{tab:1} 
\begin{tabular}{ccccc}
\hline\noalign{\smallskip}
\textbf{STARS} $ \rightarrow $ & \textbf{PSR J1903+327}  & \textbf{Vela X-1} & \textbf{Her X-1} & \textbf{SAX J1808.4-3658} \\
\noalign{\smallskip}\hline\noalign{\smallskip}
\vspace{0.2cm}R (Radius (km))                                                                & 9.438                                                                    & 9.56                                                                   & 8.1                                                                     & 7.951  \\
\vspace{0.2cm}$ M \odot $                                                                    & 1.667                                                                    & 1.77                                                                   & 0.85                                                                    & 0.9  \\
\vspace{0.2cm}L (Geometric Parameter)                                                        & 9.048                                                                    & 8.71                                                                   & 12.097                                                                  & 11.2296  \\
\vspace{0.2cm}$ p_\perp \geq 0 $                                                             & $ 0 \leq \alpha \leq 0.350528 $                                          & $ 0 \leq \alpha \leq 0.373512 $                                        & $ 0 \leq \alpha \leq 0.229171 $                                         & $ 0 \leq \alpha \leq 0.238879 $ \\
\vspace{0.2cm}$ 0 \leq \frac{dp_r}{d\rho} \leq 1$                                            & $ 0 \leq \alpha \leq 1 $                                                 & $ 0 \leq \alpha \leq 1 $                                               & $ 0 \leq \alpha \leq 1 $                                                & $ 0 \leq \alpha \leq 1 $  \\
\vspace{0.2cm}$ 0 \leq \frac{dp_\perp}{d\rho} \leq 1$                                        & $ 0.0957139 \leq \alpha \leq 0.759497 $                                  & $ 0.0967139 \leq \alpha \leq 0.749649 $                                & $ 0.087171 \leq \alpha \leq 0.690579 $                                  & $ 0.0881597 \leq \alpha \leq 0.702425 $  \\
\vspace{0.2cm}$ \rho - p_r - 2 p_\perp \geq 0 $                                              & $ -0.293786 \leq \alpha \leq 0.484878 $                                  & $ -0.246578 \leq \alpha \leq 0.468393 $                                & $ -0.79312 \leq \alpha \leq 0.737388 $                                  & $ -0.714122 \leq \alpha \leq 0.69128 $ \\
\vspace{0.2cm}$ \frac{dp_\perp}{dr} \leq 0 $                                                 & $ 0.0552885 \leq \alpha $                                                & $ 0.0505688 \leq \alpha  $                                             & $ 0.0763971 \leq \alpha  $                                              & $ 0.0753484 \leq \alpha  $  \\
\vspace{0.2cm}$ \frac{d}{dr}\left(\frac{p_\perp}{\rho}\right) \leq 0 $                       & $ 0.159617 \leq \alpha  $                                                & $ 0.166986 \leq \alpha  $                                              & $ 0.113357 \leq \alpha  $                                               & $ 0.117579 \leq \alpha  $  \\
\vspace{0.2cm}$ 0 \leq \left(\frac{dp_r}{d\rho} - \frac{dp_\perp}{d\rho} \right) \leq 1 $    & $ -1.04868 \leq \alpha \leq 0.186917 $                                   & $ -1.04722 \leq \alpha \leq 0.168048 $                                 & $ -1.03197 \leq \alpha \leq 0.219847 $                                  & $ -1.03755 \leq \alpha \leq 0.218066 $  \\
\vspace{0.2cm}$ \Gamma > \frac{4}{3} $                                                       & $ -0.121301 < \alpha $                                                   & $ -0.0718444 < \alpha $                                                & $ -0.878832 < \alpha $                                                  & $ -0.740505 < \alpha $	  \\
\vspace{0.2cm}\textbf{Final Bound on $ \alpha $}                                             & $\textbf{0.159617} \leq $ \textbf{$\alpha$} $ \leq \textbf{0.186917} $   & $\textbf{0.166986} \leq $ \textbf{$\alpha$} $ \leq \textbf{0.168048} $ & $\textbf{0.113357} \leq $ \textbf{$\alpha$} $ \leq \textbf{0.219847} $  & $\textbf{0.117579} \leq $ \textbf{$\alpha$} $ \leq \textbf{0.218066} $  \\
\vspace{0.2cm}central redshift $ (z_0) $ &  $ -1 + 1.89674 \sqrt{e^{0.492661\alpha}} $ & $ -1 + 2.00666 \sqrt{e^{0.558027 \alpha }} $ & $ -1 + 1.34622 \sqrt{e^{0.143476 \alpha }} $ & $ -1 + 1.38889 \sqrt{e^{0.170026 \alpha }} $\\
surface redshift $ (z_R) $ &  $ 0.445014 $ & $ 0.484842 $ & $ 0.203473 $ & $ 0.225284 $\\
\noalign{\smallskip}\hline
\end{tabular} 
\end{table}
\twocolumn
Thus we have physically acceptable models of superdense stars with linear equation of state and a definite 3-space geometry, viz., the paraboloidal spacetime geometry. The present model is mathematically interesting as it has got a definite geometry and physically interesting because the spacetime with linear equation of state may be good candidate for representing strange stars. \\
\bibliographystyle{plainnat} 
\bibliography{EPJAVD}

\end{document}